\documentclass[a4paper]{aa}
\usepackage{graphicx}
\usepackage{natbib}
\usepackage{amsmath}
\usepackage{amsfonts}
\usepackage{amssymb}
\usepackage{array}
\usepackage{txfonts}
\usepackage{multirow}

\begin{document}

\title{Refining the fundamental plane of accreting black holes}

\author{Elmar K\"ording \inst{1,2} \and Heino Falcke\inst{3,4} \and St\'ephane Corbel \inst{1}}

\institute{AIM - Unit\'e Mixte de Recherche CEA - CNRS - Universit\'e Paris VII  - UMR 7158, CEA-Saclay, Service d'Astrophysique, F-91191 Gif-sur-Yvette, Cedex, France
\and
School of Physics and Astronomy, University of Southampton Hampshire SO17 1BJ, United Kingdom
\and
Radio Observatory, ASTRON, Dwingeloo, P.O. Box 2, 7990 AA 
Dwingeloo, The Netherlands
         \and
           Dept. of Astronomy, Radboud Universiteit Nijmegen, Postbus 9010, 
 6500 GL Nijmegen,  The Netherlands
}

\date{2.9.2005}

\titlerunning{Refining the fundamental plane}

\abstract{The idea of a unified description of supermassive and stellar black holes 
has been supported by the extension of the empirical radio/X-ray
correlation from X-ray binaries to active galactic nuclei through the 
inclusion of a mass term. This has lead to the so-called fundamental plane of black hole activity in the black hole mass, radio and X-ray luminosity space.  Two incarnations of this fundamental plane have so far been suggested using different underlying models and using two different samples of accreting black holes.}
{We improve the parameter estimates of the fundamental plane and estimate the scatter of the sources around the plane in both samples. This is used to look for possible constraints on the proposed theoretical models. Furthermore, we search 
for selection effects due to the inclusion of different classes of AGN or distance effects.}
{We present revised samples for both studies together with a refined statistical analysis using measured errors of the observables. This method is used to compare the two samples and infer parameters for the 
fundamental plane in a homogeneous way.} 
{We show that strongly sub-Eddington objects in a state equivalent to the 
low/hard state of X-ray binaries follow the fundamental plane very tightly; 
the scatter is comparable to the measurement errors.
However, we find that the estimated parameters depend strongly on the 
assumptions made on the
sources of scatter and the relative weight of the different AGN
classes in the sample. Using only hard state objects, the fundamental plane is 
in agreement with the prediction of
a simple uncooled synchrotron/jet model for the emitted radiation. 
Inclusion of high-state objects increases the scatter and moves the 
correlation closer to a
disk/jet model. This is qualitatively consistent with a picture where
low-state objects are largely dominated by jet emission while
high-state objects have a strong contribution from an accretion
disk.}{}

\maketitle 

\section{Introduction}

Active galactic nuclei (AGN) and black hole X-ray binaries (XRBs) seem
to have a similar central engine consisting of the central black hole,
an accretion disk probably accompanied by a corona, and a relativistic
jet \citep{ShakuraSunyaev1973,MirabelRodriguez1999,Antonucci1993}.
Jet and disk may form a symbiotic system
\citep{FalckeBiermann1995,FalckeMalkanBiermann1995} which can be scaled over 
several orders of magnitude in mass and accretion rate
\citep{FalckeBiermann1996,FalckeBiermann1999} suggesting that a single central engine 
can be used to describe very different types of black holes.

While the general unification of stellar mass and supermassive black
holes picture has now been established for some time, it has recently
been tested on a detailed empirical level by correlations in the radio
and X-ray band which have lead to the so-called fundamental plane of
black hole activity (\citealt{MerloniHeinzdiMatteo2003} hereafter MHDM,
and \citealt{FalckeKoerdingMarkoff2004} hereafter FKM). Similar
unification efforts are also under way analysing and comparing the
variability properties of AGN and XRBs \citep{UttleyMcHardyPapadakis2002,MarkowitzEdelsonVaughan2003,KoerdingFalcke2004,AbramowiczKluzniakMcClintock2004}.

To establish connections between stellar and supermassive black holes
we have to consider that black hole XRBs can be found in distinct
accretion states. In FKM we suggested that a number of AGN classes can
be identified with corresponding XRB states and based on this proposed
a power unification scheme for AGN and XRBs.

The two most prevalent XRB states are the low/hard state (LH state) and
the high/soft state (HS state, see e.g.,
\citealt{McClintockRemillard2003}). In the LH state the radio spectrum
is always consistent with coming from a steady jet \citep{Fender2001}, which can
sometimes be directly imaged \citep{StirlingSpencer2001}.  Once the
source enters the HS state, the radio emission seems to be quenched
\citep{FenderCorbelTzioumis1999,CorbelFenderTzioumis2000}. One
possible scenario for the accretion flow of a LH state object is that
its inner part is optically thin up to a transition radius, where the
flow turns into a standard thin disk
\citep{EsinMcClintockNarayan1997,Poutanen1998}.  
Usually, the X-ray emission of a LH state XRB is modeled using
Comptonization (e.g. \citealt{SunyaevTruemper1979,ThornePrice1975}), however, some models suggest that the compact jet may contribute to the X-ray
emission or even dominate it
\citep{MarkoffFalckeFender2001,MarkoffNowak2004,HomanBuxtonMarkoff2005,MarkoffNowakWilms2005}. 
As the disk fades, the system may become ``jet-dominated'' ---
meaning that the bulk of the energy output is in radiation and kinetic
energy of the jet (\citealt{FenderGalloJonker2003}, FKM).

Indeed, \cite{CorbelFenderTzioumis2000,CorbelNowakFender2003} found a
surprisingly tight correlation of the radio and X-ray fluxes of the
black hole XRB GX~339$-$4 in its LH state which can be qualitatively
and quantitatively well understood in the context of jet models
\citep{MarkoffNowakCorbel2003}. \cite{GalloFenderPooley2003}
showed that this correlation does not only hold for one source but
seems to be universal for all LH state XRBs. It has also been observed
that, once an object enters the high state, the radio emission is
quenched and it drops off the correlation
\citep{FenderCorbelTzioumis1999,TananbaumGurskyKellogg1972,GalloFenderPooley2003}. It has been suggested by \citet{MaccaroneGalloFender2003} that a similar effect can be found in AGN.

Radio/X-ray correlations have also been found for AGN, e.g., by
\cite{HardcastleWorrall1999} and \cite{CanosaWorrallHardcastle1999}.  The final breakthrough came when it was
shown possible to combine these correlations to a fundamental plane in
the radio/X-ray/black hole mass space for XRBs and AGN (MHDM,
FKM). This fundamental plane gives a tight relation between the radio
and X-ray fluxes and the black hole mass, which is valid for AGN as
well as XRBs. Thus, the correlation proves the similarity of the
central engines of these accreting black holes.

However, there are at least two competing explanations for the
fundamental plane.  In the picture of FKM, the radio-through-X-ray
emission for XRBs and the lowest luminosity AGN is attributed to
synchrotron emission from a relativistic jet in the jet-dominated state
(LH). As both components, the radio and the X-rays, originate from the
same source -- the jet -- one can expect a tight correlation of both
observables. We refer to this model as the 'jet only' model. One would
expect the correlation to break down once a source leaves the
radiatively inefficient accretion flow state and is no longer jet-dominated. Hence, the picture should not apply for high-state objects. 

On the other hand, MHDM suggested that the X-ray emission originates
from the accretion flow, while the radio emission is still attributed
to the relativistic jet. Both, the flow and the jet are presumed to be
strongly coupled so that the radio and X-ray emission is
correlated. Here we will mainly assume the accretion flow to be
some variant of an advection-dominated accretion flow (ADAF,
\citealt{NarayanYi1994}, see also the convection dominated accretion flows e.g., \citealt{QuataertGruzinov2000a}) and refer to the model as 'ADAF/jet' model. The
ADAF solution is only one possible accretion flow model, e.g. one other possibility is presented in \citet{HaardtMaraschi1991}. Here we will use
the ADAF/jet model only as the example for possible 'disk/jet' models.

In the recent past, the statistics and slopes of the radio/X-ray/mass
correlations have been used to argue for and against the
synchrotron/jet models (\citealt{Heinz2004}, MHDM). Hence, further
clarification is urgently needed. Additionally,
\cite{HeinzMerloni2004} have used the correlation to search for
constrains of the relativistic beaming. However, as we will show here,
all these analyses depend strongly on the statistics of the samples,
the construction of the samples, and the assumptions on the scatter of
the measurements.

In this paper, we therefore investigate the problems of the parameter
estimation of the fundamental plane of black hole activity. We will
check the assumptions made by previous studies, and present an
improved statistical analysis. We furthermore improve the samples
presented by MHDM and FKM. With our refined parameter estimation method
we analyze and compare both samples and investigate selection effects
and the intrinsic scatter of the correlation.  In this light, we
discuss if the fundamental plane can be used to constrain the
underlying emission mechanism as previously suggested. We will use the
intrinsic scatter to test which classes of AGN belong to the analog of
the LH state XRBs.

In Sect.~2 we discuss our method of parameter estimation, the improved
samples, and discuss observing frequencies. In Sect.~3 we present our
results and their implications and present our conclusions in Sect.~4.

\section{Parameter Estimation}\label{paraestimation}
We are searching for the parameters of the fundamental plane for accreting black holes: 
\begin{equation}
\log L_{\text X} = \xi_{\text R} \log L_{\text R} + \xi_{\text M} \log M + b_{\text X}, \label{eqfunplane}
\end{equation}
where $L_{\text X}$ is the X-ray luminosity in the observed band and
$L_{\text R}$ denotes radio luminosity at the observing frequency
($\nu F_\nu$), $M$ is the black hole mass, the $\xi_{\text i}$ are
the correlation coefficients, and $b_{\text X}$ denotes the constant offset.  To simplify the notation we omit the
units in the logarithms. Throughout this paper all luminosities are
measured in erg/s, distances in pc and masses in solar masses.  In the
notation we follow FKM. To derive the parameters as given in MHDM set
$\xi_{\text RX} = 1/\xi_{\text R}$ and $\xi_{\text RM} = \xi_{\text
M}/\xi_{\text R}$.

The predicted values for the 'jet only' model are $\xi_{\text R} = 1.38$ and $\xi_{\text M} = -0.81$ (FKM), while the values for the 'ADAF/jet' model are $\xi_{\text R} = 1.64$ and $\xi_{\text M} = -1.3$ (MHDM).

\subsection{The samples}\label{secsample}
Here we compare the published correlations by MHDM and FKM.
While both samples are used to extend the radio/X-ray correlation found in LH state XRBs, they differ in the selection of sources and which observing frequencies are used for the X-ray luminosities.

\subsubsection*{MHDM sample}

The MHDM sample is a real radio/X-ray sample, i.e., it directly uses the measured radio and X-ray fluxes. It contains XRBs and nearly
all types of AGN except obviously beamed sources like BL Lac objects. The sources were extracted from the literature under the
condition that good mass estimates exist. To obtain a representative
sample the authors selected a similar number of bright active AGN and
less active AGN. The sample contains low-luminosity AGN (LLAGN),
LINERs (low ionization nuclear emission region), Seyferts (Type 1 and
2), FR Radio Galaxies \citep{FanaroffRiley1974} and radio loud and
quiet quasars \citep{KellermannSramekSchmidt1989} and the quiescence flux of Sgr A$^*$. Thus, by using
this sample one averages over nearly all types of AGN, whether they
belong to the LH state or not.

To avoid dealing with upper limits in the data we exclude all those limits from the MHDM sample. The overall result of the fit does not seem to change due to this, as we can reproduce the best fit values of MHDM. This sample contains some XRBs that do not follow the correlation. Cyg~X$-$1 changes its state frequently \citep{GalloFenderPooley2003} and seems to stay always near the transition luminosity, thus, it will not trace the correlation well.
GRS~1915$+$105 is a rather unique system that seems to stay in the 'canonical' very high state most of its time \citep{ReigBelloni2003}. We therefore exclude that source as well.
It is still open whether LS~5039 is a black hole or a neutron star binary. Furthermore, its radio spectrum is peculiar for a LH state object \citep{RiboCombiMirabel2005}. Thus, besides the original sample we will also consider the AGN subsample of MHDM and add a subsample of the \citet{GalloFenderPooley2003} XRB sample (see below). The quasar sample contains two very radio loud objects (3C273, PG~1226+023), while most quasars are radio quiet. To demonstrate the selection effects we exclude these two sources from the quasar sample when we consider subsamples of the MHDM sample.

\subsubsection*{Our sample  (KFC sample)}
The sample of FKM tries to include only objects in the LH state. FKM suggest to classify LLAGN, LINER, FR I Radio Galaxies and BL Lac objects as the analog classes of the LH state in XRBs. 
As FKM use a jet model as the basis of their suggested unification scheme, they try to compare observations at frequencies that originate from synchrotron emission (see the discussion in Sect.~\ref{SyncCut}).
They therefore extrapolate optical observations for FR I Radio Galaxies and BL Lac objects to an equivalent X-ray flux (for details see FKM). The fluxes of the BL Lac sources have been deboosted with an average Doppler factor of 7 (FKM). In the current study we further increase the number of LLAGN sources by including all sources of the \citet{NagarFalckeWilson2005} sample with $L_{\text R} < 10^{38}$ erg/s for which we found X-ray fluxes in the literature. We will refer to this augmented sample as the 'KFC sample'\footnote[1]{ Despite other associations the reader may have with this abbreviation, it is simply based on the present authorlist.}. The FKM LLAGN sample is based on X-ray observations of the LLAGN sample studied by \citet{TerashimaWilson2003}. Additionally we use fluxes from the following surveys in order of preference: The Chandra v3 pipeline (\citealt{PtakGriffiths2003},www.xassist.org), the XMM serendipitous X-ray survey \citep{BarconsCarreraWatson2002}, and the ROSAT HRI pointed catalog \citep{Rosat2000}. Finally, NGC~4258 fluxes were taken from \cite{YoungWilson2004}. 
In the KFC sample we only consider the non-Seyfert galaxies in the \citet{NagarFalckeWilson2005} sample; the Seyferts of this sample will be discussed separately as, even though they are of low luminosity, they still may belong to the supermassive analog of high state XRBs, because their black hole masses are so low. 
For Sgr A$^*$, we include the hard X-ray flare by \citet{BaganoffBautzBrandt2001a}, as the flare may be due to jet emission (see e.g., \citealt{MarkoffFalckeYuan2001}). Besides the flare we also show the result for the quiescent Sgr A$^*$ flux.

\subsubsection*{XRB sample}
The sample of LH state XRBs is based on the sample of \cite{GalloFenderPooley2003}. To avoid problems with state transitions we only include GX~334-9, V404~Cyg, 4U 1543-47, XTE 1118+480, XTE J1550-564. For all sources we only consider the data if the source is in the LH state. We excluded GRS~1915$+$105 and Cyg~X$-$1 as discussed above. For GX~334$-$9 \citep{CorbelNowakFender2003} we used the updated X-ray fluxes from \cite{NowakWilmsHeinz2005}.

\subsection{Problems of Parameter Estimation}
The correlation between the radio and X-ray emission has been discussed for XRBs and AGN before and it has been shown by partial correlation analysis that the correlation is indeed real (MHDM, for AGN only see \citealt{HardcastleWorrall1999}). For a discussion of the well constrained sources Sgr A$^*$, NGC~4258, M81 and a XRB sample see \citet{Markoff2005}. Thus, we assume that the correlation exists and only check the parameter estimation process.

To estimate parameters for measured data with errors in all variables, one has to use the merit function (see e.g., \citealt{Press2002}, MHDM). For measurements $y_{\text i}$ and  $x_{\text ij}$, which all have uncertainties, e.g., the $y_{\text i}$ can denote the X-ray luminosities while $x_{\text 1i}$ denotes the radio luminosities and $x_{\text 2j}$ the black hole masses, the merit function is defined as:
\begin{equation}
\hat{\chi}^2 = \sum_{\text i} \frac{(y_{\text i} - b - \sum_{\text j} a_{\text j} x_{\text ij})^2 }{\sigma^2_{\text y_i} + \sum_{\text j} (a_{\text j} \sigma_{\text x_{\text ij}})^2}, \label{MerritFkt}
\end{equation}
where the $\sigma$ are the corresponding uncertainties, and the $a_j$ and $b$ are the unknown parameters. 
The normal $\chi^2$ fits, which do not consider scatter in all variables, will yield asymmetric results as for them only scatter in the "y"-axis is considered. 
The derivative of the merit function is non-linear in the $a_{\text j}$  and has to be solved with a numerical optimization routine. For the parameter $b$ one can still give an analytical solution: 
\begin{equation}
b_{\text min}(\vec{a}) = \frac{\sum \omega_{\text i} \left(y_{\text i} - b_{\text j} x_{\text ij} \right)}{\sum \omega_{\text i}}
\end{equation}
with $\omega^{-1}_{\text i} = \sigma^2_{\text y_i} + \sum_{\text j} \left(a_{\text j} \sigma_{\text x_{\text ij}}\right)^2$.
Thus, we only have to solve a reasonably well behaved two dimensional function for which a standard numerical optimization routine can be used.

The resulting parameters depend strongly on the assumed uncertainties $\sigma$ in the data. MHDM assume that these uncertainties $\sigma$ are isotropic:
\begin{equation}
\sigma_{\text L_R} = \sigma_{\text L_X} = \sigma_{\text M}.
\end{equation}
Thus, they do not use measured uncertainties but set them isotropically to a value such that the reduced $\chi^2$ is unity.
This is a strong assumption and its effect has to be checked.
As a first test we explore the effect of an anisotropy of the uncertainties in the mass estimation and the scatter in the luminosities, while still assuming that the uncertainties in the $L_{\text X}$-$L_{\text R}$ plane are isotropic:
\begin{equation}
\sigma_{\text L_R} = \sigma_{\text L_X} = 2 \alpha \sigma_0 \qquad {\text and } \quad
\sigma_{\text M} = 2 (1-\alpha) \sigma_0 .
\end{equation}

\begin{figure}
\resizebox{8.7cm}{!}{\includegraphics{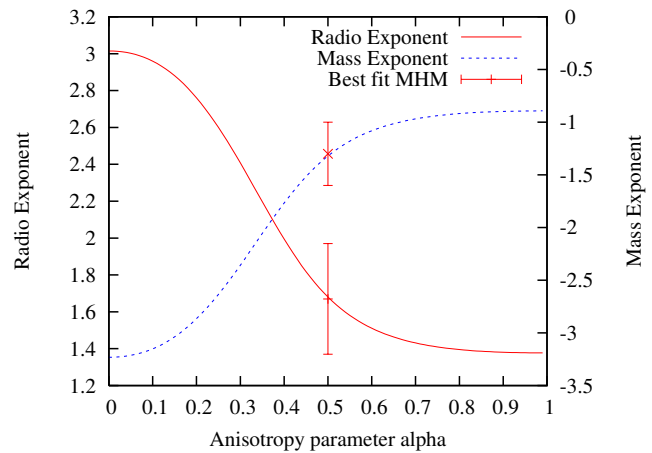}}
\caption{The effect of anisotropic errors on the best fit parameters of the fundamental plane (eq.~\ref{eqfunplane}). The horizontal axis gives the anisotropy parameter $\alpha$. A small value of $\alpha$ denotes that the uncertainties in the mass estimate dominate, while for large values the uncertainties of the luminosity are dominant. $\alpha = 0.5$ corresponds to isotropic errors as used by MHDM. Also shown are the best fit values of MHDM and their uncertainties. Depending on the choice of the uncertainties one can obtain values of the parameters $\xi_{\text R}$ between 1.4 and 3 and $\xi_{\text M}$ between -0.8 and -3.}
\label{Aniso}
\end{figure}

With these assumptions of the uncertainties, we use the merit function to derive the parameters of the original MHDM sample. The best fit values are strongly depending on the isotropy parameter  $\alpha$ as shown in Fig.~\ref{Aniso}. An anisotropy parameter of 0.5 corresponds to an isotropic distribution of the uncertainties. For this isotropic case we can reproduce the values found by MHDM, which are also shown in the figure. 
$\alpha \approx 1$ corresponds to the case that the uncertainties in the luminosities dominate while for $\alpha \approx 0$ the uncertainties of the mass estimation are dominant. We observe that, for example, the parameter $\xi_{\text R}$ can take any value between 1.4 and 3 for different $\alpha$. 

In case that the uncertainties only deviate slightly from the isotropic case the slope of the parameters tells us how strong these errors propagate to the final fit values. Unfortunately the slope of the parameters around $\alpha \approx 0.5$ is large. Thus, it is crucial to have a good estimate of the distribution of uncertainties.
If we can improve the estimates of the uncertainties we can improve the validity of the parameter estimates. Any study based on a parameter estimate using isotropic uncertainties has to take the rather large additional uncertainties due to this assumption into account.

Besides the demonstrated effect of the anisotropy of the uncertainties of the mass estimates and the luminosity estimates, a similar effect can be found if $\sigma_{\text L_R} \neq \sigma_{\text L_X}$
We note that the best fit values do not depend on the absolute value of the combined $\sigma$ but on the relative prominence of the different $\sigma_i$. 

One problem of this parameter estimation scheme is that we can not deal with coupled uncertainties. We include several measurements of the XRBs GX~339$-$4 and V404~Cyg so the uncertainties of the mass and distance measurements are not independent for these datapoint. We will neglect this effect for simplicity, but it may influence the estimated parameters especially for the small subsamples.

\subsection{Error Budget}\label{errorbudget}
Both luminosities depend on the measured flux and the distance, thus, the scatter in both quantities is coupled. To separate the equation into variables that have nearly independent errors, we separate the effect of the distance:
\begin{equation}
 L_{\text R,X} = F_{\text R,X} \Xi _{\text R,X} D^2
\end{equation}
where $F_{\text R,X}$ denote the measured radio and X-ray fluxes and $D$ denotes the distance. The $\Xi_{\text R,X}$ are conversion factors depending on the observed band.
We therefore find:
\begin{equation}
\log F_{\text X} = \xi_{\text R} \log F_{\text R} + \xi_{\text M} \log M + \left(2 \xi_{\text R} -2  \right) \log D+ b_{\text X} + \log \Xi, \label{fittedeq}
\end{equation}
where the mathematical conversion factors are combined into $\Xi$. As they are just mathematical constants, they will not be discussed further. 

We have seen in the previous section that the assumption of isotropic uncertainties (as used by MHDM) will effect the best fit parameters in an unknown way. To improve this situation, we estimate the errors attributed to each variable. The fluxes $F_{\text R,X}$ contain measurement errors and intrinsic scatter as discussed below. Besides this, the masses and distances are uncertain as well. However, as the correlation coefficient $\xi_{\text R}$ will be around 1.4 the effect of errors in the distance estimation are less severe for the correlation than for the individual luminosities as it appears with a factor $\left(2 \xi_{\text R} -2  \right)$.
As the merit function requires Gaussian errors, we are always using symmetric errors in the log-log space, many of the errors below are indeed symmetric and the other parameters are only mildly asymmetric.
\begin{itemize}
\item Mass estimate: Our XRBs LH state sample is dominated by GX~334$-$9 and V404~Cyg. For GX~339$-$4 \citet{HynesSteeghsCasares2003} estimate a mass function of $5.8 \pm 0.5$ M$_\odot$, which is therefore a lower limit for the mass of the black hole. We therefore assume a mass of $8 \pm 2.0$ M$_\odot$. For V404~Cyg we use a mass of $12 \pm 2.5$ M$_\odot$ (\citet{Orosz2003} gives a range of 10-13.4 M$_\odot$). Note that there are several data points for each object. Thus, their uncertainties are coupled, which we can not take into account. The mass estimates of the other XRBs are also taken from \citet{Orosz2003}.
For the AGN \citet{MerrittFerrarese2001} give an absolute scatter of 0.34 dex for M-$\sigma$ relation. The mass estimate using the M-$\sigma$ relation is independent of the distance of the source (cf., \citealt{FerrareseFord2005}). This method is used for all galaxies except our BL Lac objects. For masses estimated with the M-$\sigma$ relation we will use the scatter of the correlation as a measure of the uncertainty for the mass estimate: 0.34 dex. We use velocity dispersions from the Hypercat catalog \citep{PrugnielZasovBusarello1998}. For the BL Lac objects we used indirectly derived velocity dispersions from \citet{WooUrry2002}. Thus, these indirect measurements will have a higher uncertainty. \citet{BettoniFalomoFasano2001} give an uncertainty of these indirectly measured velocity dispersions $\sigma$ a value of $\delta \sigma = 18$ km/s, which yields an additional uncertainty of $\approx 0.3$ dex for the mass estimate. Thus, we use an uncertainty of 0.46 dex for the mass estimate of the BL Lac objects. The indirect method to derive the velocity dispersion depends slightly on the distance. However, as we only use this method for BL Lac objects and the distances are accurate compared to 0.46 dex uncertainty, we will ignore this effect. For Sgr A$^*$, M~81, and NGC~4258 we use direct mass measurements (see FKM), the mass uncertainties are as low as 10\% ($<0.05$ dex). 
\item Distance measurement:  The distance of GX~339$-$4 is still under debate. \citet{ShahbazFenderCharles2001} and \cite{JonkerNelemans2004} give a lower limit of 6 kpc, but the distance may be as high as 15 kpc \citep{HynesSteeghsCasares2004}. We therefore adopt a distance of $8 \pm 2$ kpc.
For V404~Cyg, we adopt the distance of $4$ kpc \citep{JonkerNelemans2004}. We use an uncertainty of 1 kpc, as we can not account for asymmetric uncertainties and have the issue of coupled errors.
For nearby AGN we use updated distances from \citet{MaozNagarFalcke2005,TonryDresslerBlakeslee2001}. If these are not available, we use the distance estimates as given in the \citet{Tully1988}. The distance uncertainty is hard to access, as many different methods are used for which the uncertainty is sometimes not well known. We assume $40 \%$, however, we checked that it does not change the result if one assumes less scatter. For AGN with distances derived from the Hubble law, we use an error estimate based on the peculiar velocities in the Hubble flow and the uncertainties of the Hubble constant (we assume H$_0$ = $72$ km/s/Mpc \citealt{SpergelVerdePeiris2003}, $5\%$ uncertainty, $\Omega_\Lambda =0.7$ and $\Omega_{\text M} = 0.3$). \citet{HawkinsMaddoxCole2003} give a peculiar velocity of 506 km/s which corresponds to 6.7 Mpc. Thus, for most sources the distance uncertainty is mainly due to the uncertainty in the Hubble flow.
For the MHDM sample, we use the distances as provided by MHDM and assume a constant uncertainty of $40\%$ in the distance estimation to avoid that we overestimate the intrinsic scatter, see below. Here, we also checked that this assumption is not critical. For most sources MHDM derive the distance from the Hubble law, so that the uncertainty is mainly due to the Hubble constant.
\item Flux measurements: For all but the faintest objects are the fluxes very well constrained. Errors for radio and optical/X-ray fluxes are usually less than 10\%. Systematic errors, e.g., due to the extrapolation of the different observed energy bands to our used X-ray band (0.5-10 keV), will also be of a similar magnitude. Thus, as these errors are small compared with the scatter due to the mass measurements, we do not introduce significant changes by assuming that these errors are isotropic.
Due to this assumption, it is possible to unify these flux errors with the intrinsic errors of the source.
\item Intrinsic errors:
Besides the measurement errors above, there are several sources of intrinsic scatter of often unknown magnitude:
\begin{itemize}
\item Non-simultaneous observations of the AGN: The radio and X-ray observations are non-simultaneous, there is often more than a year between the different observations. All accreting black holes (AGN and XRBs) are highly variable. Thus, already this effect can lead to a deviation by more than an order of magnitude. The orientation of this uncertainty is likely to be isotropic.
\item Beaming: In most models at least the radio emission is attributed to the relativistic jet and will thus be relativistically beamed. In case that the X-rays originate from the disk/corona they will not be beamed and the deviations from the correlation will be enormous. For jet models, the X-ray emission may be beamed like the radio emission or have a different beaming patters (e.g., a velocity structure in the jet \citealt{ChiabergeCelottiCapetti2000,TrussoniCapettiCelotti2003}). The asymmetry of this effect depends on the exact model, so we can only assume isotropy. 
\item Source peculiarities: The surrounding environment of the black hole will play a role on the exact emission properties (e.g., there might be compact hotspots). There may also be an obscuring torus or other obstacles for the emission. This can result in strong X-ray absorption or the radio emission may also be absorbed. All models, however, only consider the nuclear emission.
\item Spectral energy distribution (SED):
Depending on the real emission model, it may be that we are not observing the same emission type in the X-rays for the different objects. For jet models the effect of radiative cooling and the synchrotron cut-off have to be mentioned (see Sect. \ref{SyncCut}). For disk models a similar effect may be due to the relative strength of the disk component, the jet component as observed in the resolved X-ray jets, and the Comptonization component. The X-rays in XRBs may not originate from the same process as those in AGN due to the mass scaling by whatever theory is used. 
\end{itemize}
The total intrinsic scatter will be derived from the scatter of the correlation. We will see below that the intrinsic scatter is surprisingly small considering this long list of possible errors.
\end{itemize}
For the two last sources of scatter, the flux measurements and the intrinsic errors, we do not have an exact knowledge of their magnitude and their asymmetry. We will therefore assume that they are isotropic in the $\log F_{\text X}$ - $\log F_{\text R}$ plane and parameterize their combined magnitude as $\sigma_{\text Int}$. This parameter $\sigma_{\text Int}$ will be chosen, such that the reduced merit function is unity.  With this choice we assume the error distribution just described; this will therefore affect the final fit values. If one distributes the excess variance in a different manner one will find slightly different best fit parameters. However, as we know several effects that introduce uncertainties in the radio and X-ray fluxes and the the excess variance in these variables is surprisingly low, it is sensible to include the excess variance only in the radio and X-ray fluxes.

\subsection{The parameter $\sigma_{\text Int}$}
In Sect.~\ref{errorbudget} we present our assumptions on the uncertainties of the measured variables. In case that these assumptions are exact, e.g., the magnitude of the uncertainties and that they are Gaussian distributed, the fitted parameter $\sigma_{\text Int}$ will describe the real intrinsic scatter of the sources. This value could then be used to constrain the different contributions like the effect of beaming. 
However, if we overestimate the uncertainties in the flux, distance, and mass estimation the derived $\sigma_{\text Int}$ will be too small. Similarly, an underestimation will lead to an overestimation of the intrinsic scatter.

Our method can not treat coupled uncertainties correctly. For XRBs we include several data points for each source, but there is only one mass and distance estimates for that source. Thus, as we have to assume that our uncertainties are independent, we overestimate the measurement errors.
Thus, the inferred $\sigma_{\text Int}$ for XRBs alone is zero. This means that the deviations of the data points from the optimal correlation is within the measurement uncertainties. On the other hand, for the AGN samples the uncertainties are independent. Every source is included only once in the sample. Here, $\sigma_{\text Int}$ should be a good measure for the intrinsic scatter.

\subsection{Origin of X-ray emission in the jet model} \label{SyncCut}
\begin{figure}
\resizebox{10cm}{!}{\includegraphics[angle=0]{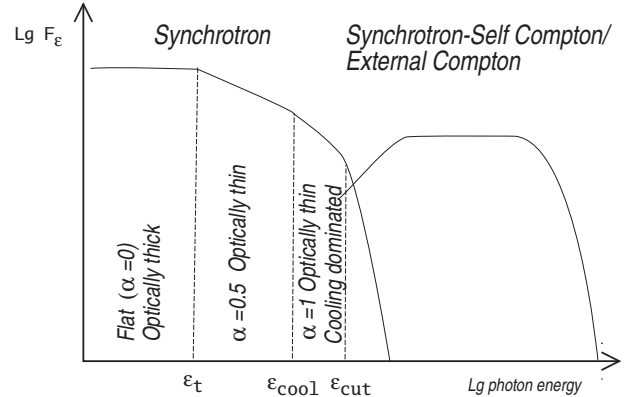}}
\caption{Sketch of the SED of a relativistic jet as for example observed in BL Lac objects. At lower photon energies synchrotron emission is dominant while the high energy spectrum is usually explained by inverse Compton processes. See text for details. }
\label{fiJetSpec}
\end{figure}

The spectrum of a relativistic jet can be directly observed in BL Lac objects, as relativistic boosting increases the relative prominence of the jet component compared to the disk \citep{BlandfordRees1978}. An idealized spectrum of a jet, i.e., the "Camel's back", is shown in Fig.~\ref{fiJetSpec} in flux ($F_{\epsilon}$) representation. Such a jet component exists at least in every AGN with a detectable jet, most likely in all AGN. The relative prominence of this component in respect to disk and corona emission will vary. Furthermore, the exact shape of the SED depends also on the inclination angle, the Lorentz factor of the jet and peculiarities of the source. 

The spectrum of a conical jet is flat (in $F_\nu$ representation) due to optically thick synchrotron emission up to the turnover frequency $\epsilon_{\text t}$. This is followed by the optically thin power law component with a typical energy index $\alpha$ $(F_{\epsilon} \sim \epsilon^{-\alpha})$ of around $\alpha = 0.5$. This emission comes from the innermost region of the radiating jet. The power law continues up to the energy where radiative cooling plays a role ($\epsilon_{\text cool}$); here the power law may steepen to $\alpha \approx 1 $. Finally the synchrotron emission cuts off at ($\epsilon_{\text cut}$), due to the acceleration mechanism creating the radiating particles. At photon energies above the synchrotron cut-off, synchrotron-self Compton emission and external Compton emission are visible. As the jet is relativistically boosted, the total observed power in these two humps depend on the jet Lorentz factor and the inclination of the source. The relative prominence of the synchrotron and the inverse-Compton emission depends on these two parameters as well.  

The photon energy where the synchrotron emission cuts off, seems to depend on the total power of the accreting system, see e.g., the blazar sequence \citep{FossatiMaraschiCelotti1998,GhiselliniCelottiCostamante2002}.
For low power systems, like high-peaked BL Lac objects (see the previous references), the cut-off energy can be above or around the standard X-ray band of 0.5-10 keV, while for strongly accreting systems, e.g., flat spectrum radio quasars, this cut-off can be as low as 0.1 eV. Thus, we can expect that only the X-rays of  low power systems actually originate from synchrotron emission.

Radiative cooling in the X-rays, as for example discussed by \cite{Heinz2004}, may play a role in some sources of low to intermediate accretion rates. Its main effect will be that the measured X-ray flux will be reduced to what one would expect from a simple uncooled jet. However, the X-ray reduction due to cooling is far less severe than the synchrotron cut-off. The effect of cooling will increase the observed intrinsic scatter as we are exploring only the simplest jet model: the uncooled jet. For most AGN besides the LLAGN in the MHDM sample this synchrotron cut-off will be below the X-ray band. The X-ray emission in these sources will not be due to synchrotron emission. It will either originate from inverse-Compton processes in the jet or from the disk/corona.

The cut-off energy ($\epsilon_{\text cut}$) of LLAGN sources in the KFC sample will for most sources be above the ROSAT and Chandra bands. The radio luminosities of the LLAGN are in the range of $10^{36-38}$ erg/s. If one extrapolates Fig.~7 of \citet{FossatiMaraschiCelotti1998} to these energies, one can expect a cut-off above $10^{19-20}$ Hz or 40-400 keV. For BL Lac objects and FR-I RGs, the KFC sample extrapolates optical fluxes to an equivalent X-ray flux. This treatment avoids the effect of the synchrotron cut-off also for these sources. The 'jet only' model should therefore be applicable to the KFC sample. 

\subsection{Origin of the X-ray emission in the MHDM sample}\label{XRayMHDM}
The X-ray emission of many sources in the MHDM sample is very likely not due to synchrotron emission, due to the synchrotron cut-off. For example, for radio loud quasars, the synchrotron cut-off is far below the X-ray regime (see e.g., \citealt{TavecchioMaraschiGhisellini2002}). Only the X-ray emission from the LLAGN may originate from synchrotron emission. Thus, the 'jet-only' model is not applicable to the MHDM sample, and can therefore not be constrained using the MHDM sample.

The AGN sample of MHDM contains radio loud objects (e.g., Cyg~A, 3C273, etc). The X-ray spectral index of radio loud Quasars (RLQ) and those of radio quiet Quasars (RQQs) seems to be different: $\Gamma_{\text RLQ} \approx 1.63\pm 0.02$ compared to $\Gamma_{\text RQQ} \approx 1.89\pm0.11$. Furthermore, RLQ show very weak or no reflection components and the strength of the soft excess seems to be anti-correlated with the radio-loudness. These effects are usually explained by a relativistic jet component in the RLQ case (see e.g., \citealt{ReevesTurner2000,PiconcelliGuainazzi2005}). 

On the other hand, there are several detected "resolved" X-ray jets in AGN. For the source 3C273, which is included in the MHDM sample, see \cite{MarshallHarrisGrimes2001}. \citet{MarshallSchwartzLovell2005} and \citet{SambrunaGambillMaraschi2004} find in more than 50\% of the observed radio loud sources resolved X-ray jets. The authors suggest that all bright radio jets may have X-ray counterparts. The X-ray emission from the AGN jets is usually explained by non-thermal processes (synchrotron, inverse-Compton, synchrotron-self-Compton), which process dominates seems to vary from source to source. For a discussion see \citet{HarrisKrawczynski2002}.
As the jet is visible, it most likely contributes at least at some level to the core X-ray flux. Furthermore, many of the MHDM fluxes are derived from ASCA or GINGA data. In that case, the Chandra resolved X-ray jet will be observed as a point source. 
Note that the X-ray jets in AGN are often dominated by emission from knots, unlike what is expected for LH state XRBs as modeled by \citet{MarkoffFalckeFender2001}.
From the statistical studies and the direct observation of X-ray jets, we conclude that {\it at least some radio loud AGN in the MHDM sample have an X-ray component originating from the jet.}

Besides jet components, other features can contaminate the X-ray fluxes, especially for non-Chandra data. The flux may consist of several components originating from different physical processes, including the disk described by any model, the corona, the reflection component, warm gas and the jet (at least in radio loud objects). 

\subsection{Uncertainties of the estimated parameters}
Given the data points and their estimated uncertainties we can derive the optimal fit parameters from eq.~\ref{MerritFkt}. The merit function $\hat{\chi}^2$ can be used similarly to the usual $\chi^2$ to estimate the confidence region of the parameters. The 1 $\sigma$ confidence region should be given by $\Delta\hat{\chi}^2 \approx 2.3$ as we have 2 degrees of freedom in our model besides the offset $b_{\text X}$. As problems may arise due to the use of the nonlinear merit function or the unknown distribution function of the errors, we checked that this confidence region is in agreement with the confidence region derived by a Monte Carlo simulation and the Bootstrap method (see below).

In the Monte-Carlo simulation, the errors of the parameters are estimated by creating a large number (5000) of artificial datasets that have similar statistical properties compared to the measured dataset. For each of the artificial datasets, we estimate the best-fit parameters using the same method as for the original dataset. From these fitted parameters, we derive the confidence region and the $\Delta\hat{\chi}^2$ corresponding to the 1, 2 and 3 $\sigma$ confidence regions. 
To create the artificial data, we consider each population of sources of our measured dataset individually and measure their scatter and position in the $\log L_{\text R}$ -- $\log L_{\text X}$ plane. In the artificial dataset, we distribute the sources uniformly over their radio luminosities.
The artificial dataset will therefore also contain simulated objects of all considered types. We assume that the scatter compared to the real correlation is Gaussian. As seen in Fig.~\ref{NormalError} this seems to be roughly the case. We also introduce flux limits in the simulation to access the effects of the distance selection effect.

The bootstrap method functions as follows. From a set of $N$ measured
sources, we draw $N$ at random with replacement, thus, creating an artificial dataset. This dataset contains
some of the sources more than once, while others are omitted. The
parameters of the fundamental plane will be estimated for this sample
in the same way as for the original dataset. These simulated
parameters should be distributed around the original best fit values
as the measured parameters are distributed around the real parameters
\citep{Press2002}. The benefit of this method is that it does not
require prior knowledge of the distribution function from which the
original dataset was drawn.

\subsection{ Different parameter estimators}
\begin{table*}
\label{estimators}
\caption{Performance of the different parameter estimators. The given uncertainties is the standard deviation of the fit parameters obtained from different artificial datasets.} 
\begin{center}
\begin{tabular}{lcccc}
\hline \hline
\noalign{\smallskip}
Estimator &  $\xi_{\text R}$ & $\xi_{\text M}$ & $b_{\text X}$ \\
\hline \noalign{\smallskip}
Assumed Parameters & 1.40 & -0.85 & -4.9 \\
Merrit Function & $1.42 \pm 0.13$ & $-0.87 \pm 0.16$ & $-5.4 \pm 3.6$ \\
Merrit Function with exact knowledge of scatter & $1.40\pm 0.08$ & $-0.85\pm 0.09$ & $-4.8\pm 2.3$ \\ 
Merrit Function with isotropic uncertainties & $1.54 \pm 0.16$ & $-1.03 \pm 0.19$ & $-8.5 \pm 4.4$ \\
Maximum Likelihood & $1.10 \pm 0.08$ & $-0.47 \pm 0.09$ & $3.5 \pm 2.1$\\
Maximum Likelihood with exact knowledge of scatter &  $1.25 \pm 0.06$ & $-0.68 \pm 0.07$ & $-0.51 \pm 1.68$ \\
Ordinary least squares &  $1.09 \pm 0.08$ & $-0.46 \pm 0.09$ & $4.0 \pm 2.3$ \\
\end{tabular}
\end{center}
\end{table*}

Up to now we have only discussed the parameter estimation using the Merrit function (eq.~\ref{MerritFkt}). 
This method minimizes the average distance of the data-points from the plane weighted with the measured uncertainties.
Other possible methods used in astronomy include the ordinary least squares estimation or Maximum Likelihood methods (see e.g., \citealt{DAgostini2005}). Even though Maximum Likelihood estimators find the most probable parameters of a model, the method is often biased towards lower fit parameters. To find the optimal fitting method for our problem, we compare the different fitting methods using a Monte Carlo simulation.
We create several artificial samples with our Monte Carlo simulation and compare the results of the different estimators with the parameters used to create the sample. We set the intrinsic scatter of our artificial XRB and LLAGN sample to 0.2 and for BL Lac and FR-I RGs to 0.8, which is roughly double of what is found in our sample. The resulting intrinsic scatter of the simulated sample is $\sigma_{int} = 0.65$. For each parameter estimator consider two cases: First, we only use the average intrinsic scatter (0.65) to estimate the parameters, and second, we use the exact probability distribution used to create each individual data-point. 

For each estimator we simulated 100 different datasets and give the average estimated parameters and the standard deviation in table \ref{estimators}. While the Merrit function seems to be a fairly robust method, the maximum likelihood estimator is biased towards smaller fit values. Thus, we will estimate our parameters with the method using the Merrit function described in section \ref{paraestimation}.

\begin{table*}
\begin{center}
\caption{Best fit results for the KFC sample and the original and edited MHDM sample. Besides the full sample we also give the parameters for the subsamples containing only a limited set of AGN classes. Note that LINER sources classified by MHDM are not limited to LLAGN. The LH XRB sample is defined in Sect.~\ref{secsample}. The column N denotes the number of sources in the sample.\label{tabresult}
 } 
\begin{tabular}{lcccccccc}
\hline \hline
\noalign{\smallskip}
&  $\xi_{\text R}$ & $\xi_{\text M}$ & $b_{\text X}$& $\sigma_{\text int}$ &$\xi_{m}|\xi{r} = 1.4$ & $\sigma_{int}|\xi{r} = 1.4$ & $N$ \\
\hline \noalign{\smallskip}
\multicolumn{2}{l}{KFC sample}\\
\hline
\noalign{\smallskip}
Full Sample & $1.41 \pm 0.11$ & $-0.87 \pm 0.14$ & $-5.01 \pm 3.20$ & 0.38 & $-0.86 \pm 0.02$ & 0.38 & 77\\
(XRB, Sgr A$^*$, LLAGN, FR-I, BL Lac) \\
\noalign{\smallskip}Full Sample + Seyferts \& Transition obj. & $1.48 \pm 0.13$ & $-0.95 \pm 0.16$ & $-6.89 \pm 3.83$ & 0.44 & $-0.86 \pm 0.02$ & 0.44 & 100\\
\noalign{\smallskip}XRB, Sgr A$^*$, LLAGN, FR-I& $1.25 \pm 0.10$ & $-0.74 \pm 0.12$ & $-0.46 \pm 2.93$ & 0.28 & $-0.91 \pm 0.02$ & 0.30 & 58\\
\noalign{\smallskip}XRB, Sgr A$^*$, LLAGN, BL Lac, & $1.64 \pm 0.13$ & $-1.08 \pm 0.14$ & $-11.67 \pm 3.59$ & 0.18 & $-0.81 \pm 0.02$ & 0.23 & 62\\
\noalign{\smallskip}XRB, Sgr A$^*$, LLAGN & $1.59 \pm 0.21$ & $-1.02 \pm 0.21$ & $-10.15 \pm 6.17$ & 0.12 & $-0.84 \pm 0.02$ & 0.15 & 43\\
\noalign{\smallskip}XRB, Sgr A$^*$, LLAGN, Seyfert \& Transition &  $1.86 \pm 0.35$ & $-1.33 \pm 0.36$ & $-17.98 \pm 9.92$ & 0.35 & $-0.85 \pm 0.02$ & 0.39 & 66\\
\noalign{\smallskip}Sgr A$^*$, LLAGN, FR-I, BL Lac &  $1.70 \pm 1.17$ & $-2.17 \pm 4.09$ & $-5.21 \pm 12.08$ & 0.45 & $-1.15 \pm 0.38$ & 0.46 & 52\\
\noalign{\smallskip}\hline\noalign{\smallskip}
\noalign{\smallskip} Full Sample with quiescent Sgr A$^*$ &  $1.52 \pm 0.14$ & $-1.00 \pm 0.18$ & $-8.15 \pm 4.11$ & 0.39 & $-0.85 \pm 0.02$ & 0.40 & 77 \\
\noalign{\smallskip}XRB, quiescent Sgr A$^*$, LLAGN & $1.97 \pm 0.11$ & $-1.41 \pm 0.10$ & $-21.05 \pm 3.05$ & 0.00 & $-0.86 \pm 0.02$ & 0.23 & 43\\
\noalign{\smallskip}\hline\noalign{\smallskip}
Original MHDM sample &$1.45 \pm 0.17$ & $-0.99 \pm 0.22$ & $-5.98 \pm 5.02$ & 0.72 & $-0.93 \pm 0.03$ & 0.73 & 116\\  
\noalign{\smallskip}\hline \noalign{\smallskip}
\multicolumn{3}{l}{MHDM Sample with LH XRBs}\\
\hline \noalign{\smallskip}
Full Sample & $1.74 \pm 0.20$ & $-1.35 \pm 0.27$ & $-14.23 \pm 5.75$ & 0.65 & $-0.92 \pm 0.03$ & 0.68 & 103\\
(XRB, Sgr A$^*$, LINER, Quasar, Seyfert) \\
\noalign{\smallskip}XRB, Sgr A$^*$, LINER, Seyfert &  $1.55 \pm 0.19$ & $-1.15 \pm 0.24$ & $-8.93 \pm 5.37$ & 0.64 & $-0.96 \pm 0.03$ & 0.64 & 92\\
\noalign{\smallskip}XRB, Sgr A$^*$, LINER, Quasar &  $2.12 \pm 0.31$ & $-1.75 \pm 0.38$ & $-25.20 \pm 8.81$ & 0.51 & $-0.90 \pm 0.04$ & 0.62 & 57\\
\noalign{\smallskip}XRB, Sgr A$^*$, Quasar, Seyfert&  $1.79 \pm 0.28$ & $-1.39 \pm 0.37$ & $-15.68 \pm 7.87$ & 0.63 & $-0.88 \pm 0.03$ & 0.65 & 82\\
\noalign{\smallskip}XRB, Sgr A$^*$, LINER &  $1.59 \pm 0.35$ & $-1.19 \pm 0.40$ & $-10.11 \pm 10.13$ & 0.50 & $-1.54 \pm 0.21$ & 0.71 & 46\\
\noalign{\smallskip}XRB, Sgr A$^*$, Quasar &  $2.03 \pm 0.28$ & $-1.55 \pm 0.36$ & $-22.76 \pm 7.86$ & 0.14 & $-0.74 \pm 0.04$ & 0.32 & 37\\
\noalign{\smallskip}Sgr A$^*$, LINER, Quasar, Seyfert & $1.65 \pm 0.26$ & $-1.72 \pm 0.50$ & $-8.02 \pm 8.06$ & 0.74 & $-1.41 \pm 0.22$ & 0.76 & 78\\
\noalign{\smallskip}\hline
\end{tabular}
\end{center}
\end{table*}

\section{Results}
\begin{figure}
\resizebox{8.7cm}{!}{\includegraphics[angle=0]{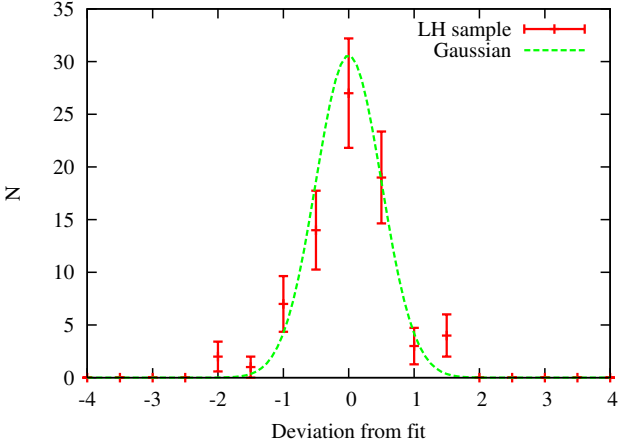}}
\caption{Histogram of the scatter in the full KFC sample. The scatter can be well approximated by a Gaussian with the standard-deviation $\sigma = 0.50$.}
\label{NormalError}
\end{figure}

Our fitting algorithm is only robust if our sources are normally distributed around the fundamental plane.
As a first test we show a histogram of the scatter of the KFC sample around the best fit to the fundamental plane in Fig.~\ref{NormalError}. The deviations are roughly normally distributed with $\sigma \approx 0.5$. Thus, the developed analysis method can be utilized.
 
\begin{figure*}
\begin{center}
\resizebox{6.7cm}{!}{\includegraphics[angle=0]{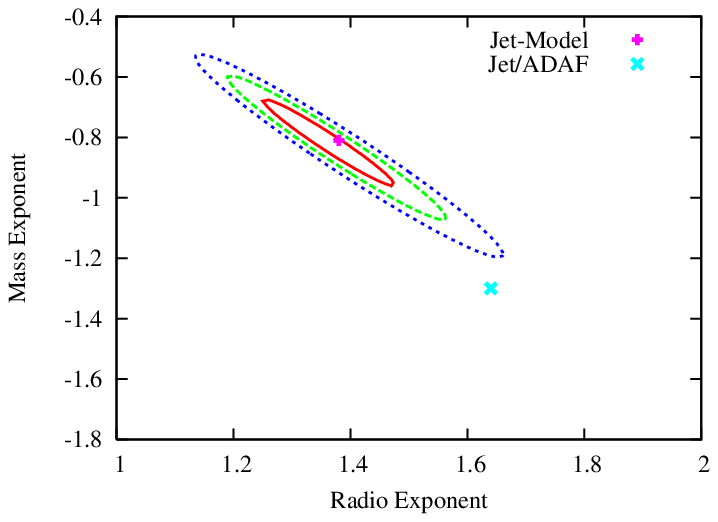}}
\resizebox{6.7cm}{!}{\includegraphics[angle=0]{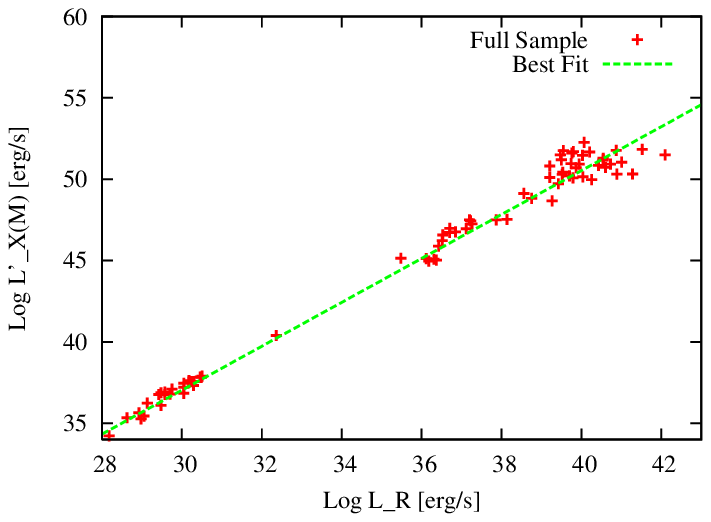}}
\resizebox{6.7cm}{!}{\includegraphics[angle=0]{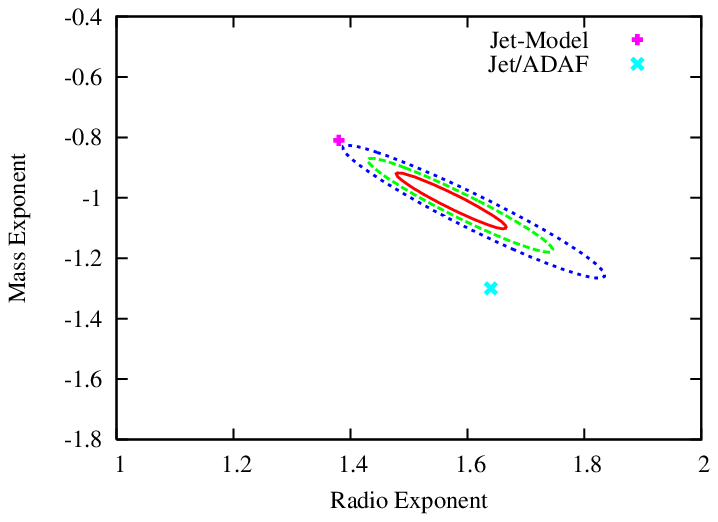}}
\resizebox{6.7cm}{!}{\includegraphics[angle=0]{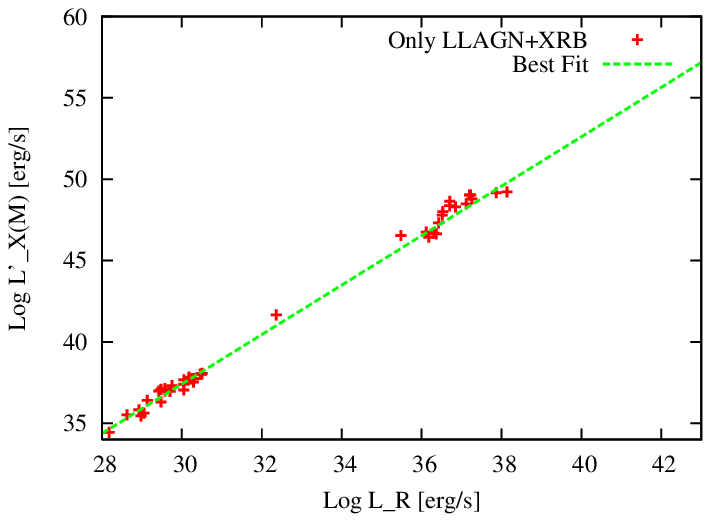}}
\resizebox{6.7cm}{!}{\includegraphics[angle=0]{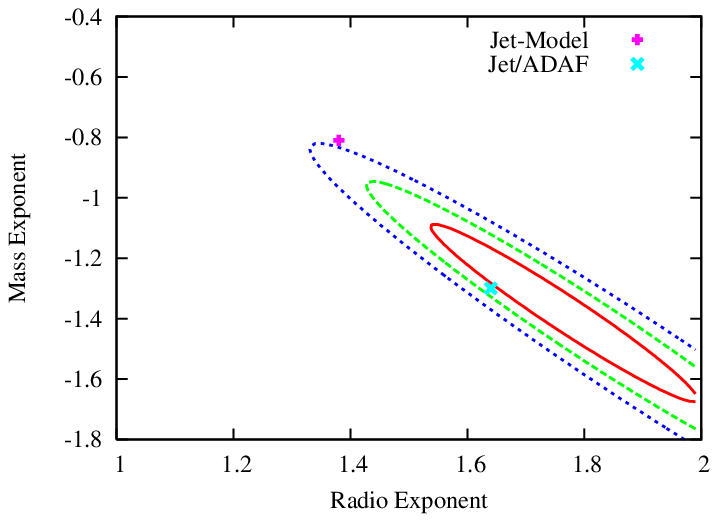}}
\resizebox{6.7cm}{!}{\includegraphics[angle=0]{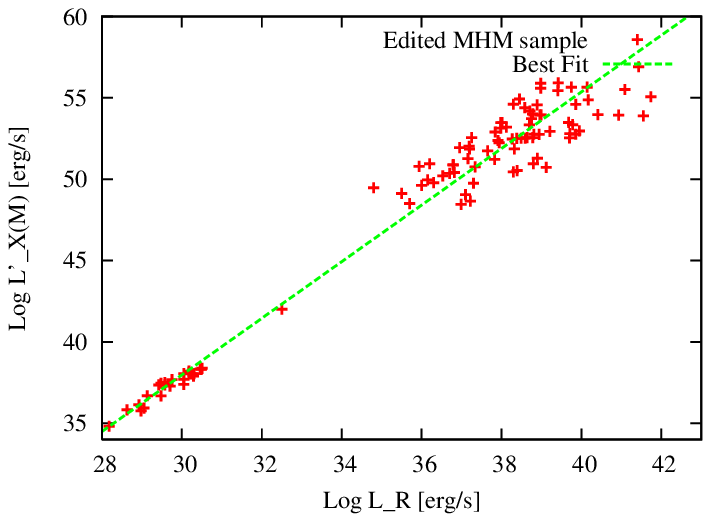}}
\end{center}
\caption{On the left side we show the $\chi^2$ map with one (solid line), two (dashed) and three sigma (dotted) levels together with the predictions of the 'jet only' model and the ADAF/jet model. The right side shows the best fit and the AGN and XRB sample. The three rows show the different samples, from Top to bottom: KFC sample, Only XRBS and LLAGN of the KFC sample, and the edited MHDM sample.}
\label{Chi2map}
\end{figure*}

We fitted the KFC sample with the fundamental plane described in eq.~\ref{fittedeq}. The parameter estimation method is described in Sect.~\ref{paraestimation}. We find as best fit values
\begin{equation}
 \xi_{\text R} = 1.41_{-0.12}^{+0.14}\quad 
\xi_{\text M} = -0.87_{-0.17}^{+0.15}\quad
b_{\text X} = -5.01_{-3.9}^{+3.35}.
\end{equation}
The confidence region of the two relevant parameters and the fit is shown in the top panel of Fig.~\ref{Chi2map}. The given uncertainties are derived from the $\Delta\hat{\chi}^2$ map. 
The intrinsic scatter in the fundamental plane is for this sample $\sigma_{\text int} = 0.38 \pm 0.06$. The uncertainty of $\sigma_{\text int}$ has been derived by bootstrapping the measured sample.

The results of the KFC sample and the MHDM sample and their subsamples is presented in Table~\ref{tabresult}. The errors given in this table are derived using the Bootstrap method. As the errorbars for the KFC sample are nearly identical for both methods, we present only the bootstrapped errorbars for the different subsamples. This error estimation is less dependent on the assumption of normal distributed scatter.  As the uncertainties are not strongly anisotropic, we give a symmetric uncertainty to avoid problems with anisotropic errors. We note that the different fit values do not coincide within the errors for both samples due to the different source populations included in the sample. 

The subsample of the KFC sample containing no FR-I radio galaxies has a larger radio coefficient ($\xi_{\text R}=1.64 \pm 0.13$) and a smaller mass coefficient ($\xi_{\text M}=-1.08 \pm 0.14$) than the full sample. On the other hand, the subsample without BL Lac objects deviates in the other direction ($\xi_{\text R}=1.25 \pm 0.1$ and $\xi_{\text M}=-0.74 \pm 0.12$). This difference arises, because FR-I radio galaxies are fainter in the optical and X-rays than the deboosted BL Lac objects \citep{ChiabergeCelottiCapetti2000}. The difference in the optical/X-ray luminosity can be explained by a velocity structure in the jet. 
As we include a similar number of BL Lac objects and FR-I radio galaxies in the full sample, these effects will partly average out. However, the final value for the parameters will depend on the total weight each AGN class has in the total sample.

The KFC subsample containing only XRBS, Sgr A$^*$ and LLAGN has similar best fit values as the subsample containing no FR-I RGs. Its correlation coefficient, $\xi_{\text R} = 1.59 \pm 0.21$, is larger than the one found for the full sample. This is partly due to the fact that the intrinsic scatter of this subsample is extremely low:  $\sigma_{\text int} = 0.12$. Thus, the errors of the mass estimation dominate (0.34 dex) the overall error budget and one gets larger fit values as shown in Fig.~\ref{Aniso}. 
If one adds Seyferts and Transition objects, which may correspond to the high state, the correlation coefficient gets even larger. 
However, all KFC subsamples seem to be roughly in agreement with  $\xi_{\text R} \approx 1.4$ and $\xi_{\text M} \approx -0.8$. This radio coefficient is in agreement with the radio/X-ray correlation for GX~339$-$4, which has $\xi_{\text R} \approx 1.4$ \citep{CorbelNowakFender2003}.

If one uses the quiescent flux of Sgr A$^*$ instead of the flare by \citet{BaganoffBautzBrandt2001a} in the KFC sample the best fit values for $\xi_{\text R}$ increase. The subsample containing only XRBs, the quiescent Sgr A$^*$ and LLAGN yields  $\xi_{\text R} = 1.97$ and  $\xi_{\text M} \approx -1.41$. Interestingly, the latter subsample has an intrinsic scatter of zero, i.e., the scatter of the correlation is in agreement with the assumed measurement errors. However, the inferred fit for $\xi_{\text R}$ is no longer in agreement with the value found for XRBs only ($\xi_{\text R} = 1.4$). 
The quiescent X-ray emission from Sgr A$^*$ is an extended \citep{BaganoffMaedaMorris2003}, while this is not the case for cores of the XRBs or AGN. Thus, it is not surprising that the fit values change. 

 For the previous fits we have used the same intrinsic scatter $\sigma_{\text int}$ for all our objects. However, we have seen that XRBs and LLAGN can be fitted with significantly less excess scatter than the full sample. Thus, we can fix the intrinsic scatter of XRBs and LLAGN to 0.1 dex and fit only the scatter in the BL Lac and FR-I RGs. Now, we find
$\xi_{\text R} = 1.61 \pm 0.11$ and  $\xi_{\text M} = -1.08$ which is similar to the result found for XRBs, Sgr A$^*$ and LLAGN only. The intrinsic scatter found for FR-I RGs and BL Lac objects is 0.55 dex. The statistical weight for these two classes is therefore significantly less than that of the LLAGN and XRBs. It is therefore not surprising that we find similar fit values: the fitting method puts only very a low weight on the additional sources. To observe selection effects it is therefore sensible to use a constant intrinsic scatter for all sources.

For the full edited MHDM sample, we find $\xi_{\text R} \approx 1.74_{-0.19}^{+0.23}$ and $\xi_{\text M} \approx -1.35_{-0.30}^{+0.24}$. As for the KFC sample, we also find a selection effect for the MHDM sample. The MHDM subsample containing no Seyfert objects yields as best fit parameters $\xi_{\text R} \approx 2.12$ and $\xi_{\text M} \approx -1.75$, while the one containing no Quasars gives $\xi_{\text R} \approx 1.55$ and $\xi_{\text M} \approx -1.15$. This effect can be explained by the fact that the quasars are more radio quiet than the other AGN in the sample.
We note that the formal fits of $\xi_{\text R}$ for most subsamples tend to be significantly higher than the value found by  \citet{CorbelNowakFender2003} of $\xi_{\text R} = 1.4$ for XRBs.
Overall, we find the choice of the sample strongly influences the final fit value in both considered samples (MHDM and KFC). This is not necessarily worrisome, as one does expect somewhat different results for different black hole states.

\begin{figure}
\resizebox{8.7cm}{!}{\includegraphics[angle=0]{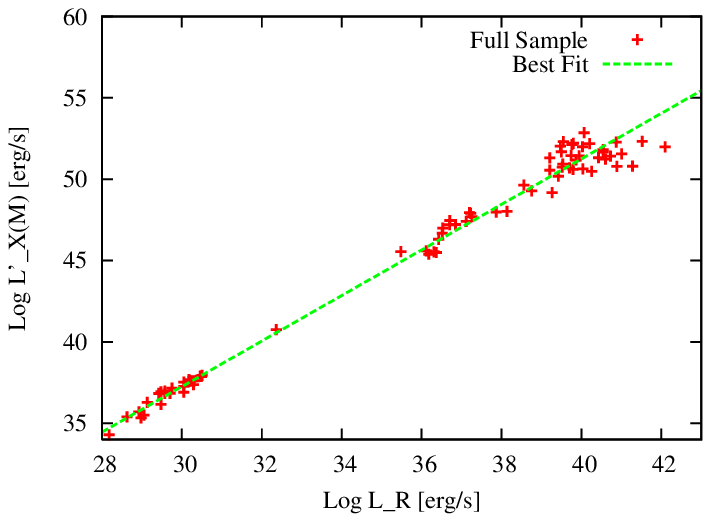}}
\caption{Best fit for the full KFC sample if one fixes $\xi_{\text R}$ to $1.4$, which is the value found for XRBs. Only the mass scaling parameter is fitted. One finds: $\xi_{\text M} = -0.86\pm 0.02$ and $b_{\text X} = - 4.9$. The intrinsic scatter increases by less than $0.01$ compared to the fit in both parameters $(\sigma_{\text int} = 0.38)$.}
\label{SetXiR}
\end{figure}

The correlation index  $\xi_{\text R}$ for XRBs only is better constrained than in the case we are considering here, as one does not have to consider the mass scaling in that case. For XRBs only one finds $\xi_{\text R} \approx 1.4$ \citep{CorbelNowakFender2003, GalloFenderPooley2003}. If we use this prior knowledge to fix the correlation index  $\xi_{\text R}$ to $1.4$ and only fit the mass scaling parameter  $\xi_{\text M}$ to our full KFC sample, we find $\xi_{\text M} = -0.86\pm 0.02$ (see table \ref{tabresult}). This fit is shown in Fig.~\ref{SetXiR}. The fit is not significantly worse than the original fit in both parameters, as the intrinsic scatter increases by less than $0.01$ $(\sigma_{\text int} = 0.38)$. 
Interestingly, the best fit value for the subsample of LLAGN, that had a higher correlation ($\xi_{\text R} =1.59$) yields with the prior knowledge $\xi_{\text M} = -0.84 \pm 0.02$, in agreement with the value found for the full sample. For the MHDM sample, we find $\xi_{\text M} = -0.92\pm 0.04$. As the correlation index $\xi_{\text R}$ is well constrained for XRBs, fixing $\xi_{\text R} = 1.4$ will likely yield the best estimates for the parameters of the fundamental plane.

Besides checking the effect of anisotropic errors we also explore the effect that the uncertainties are not normally distributed but exponentially. In that case, one does not minimize the square of the deviations divided by the uncertainty but the absolute value. This "robust" estimation \citep{Press2002} yield $\xi_{\text R} = 1.39$ and  
$\xi_{\text M} = -0.84$. For the edited MHDM sample we find $\xi_{\text R} = 1.63$ and  $\xi_{\text M} = -1.14$. Thus, the change of the distribution of the uncertainties does not significantly change the best fits. 

In table \ref{tabresult}, we also show the intrinsic scatter measured for the different datasets. If we drop our assumption that the intrinsic scatter is the same for AGN and XRBs, e.g. if $\sigma_{\text int}$ is smaller for XRBs than for AGN, then the correlation coefficient $\xi_{\text R}$ will rise slightly.

\subsection{ The distance selection effect}
Our sample of AGN and XRBs is a random sample of well studied objects and not a complete distance limited sample. Thus, one can fear that the fundamental plane is a spurious correlation created by distances selection effects. It has been shown with partial correlation analysis by MHDM that the fundamental plane is real and not spurious. This problem has been discussed in detail by \citet{MerloniKoerdingHeinz2006}, where the authors show further statistical and observational evidence that the fundamental plane is indeed real. Here, we will just present some further tests for our sample. 

The fundamental plane is an extension of radio/X-ray correlations in XRBs \citep{GalloFenderPooley2003}. This correlation is highly significant and is not effected by distance effects as we can trace individual objects on the correlation.
In AGN, \citet{HardcastleWorrall1999} and \citet{CanosaWorrallHardcastle1999} showed that a radio/X-ray correlation exists and is also not created as an artifact of plotting "distance against distance" in a flux limited sample. If one combines these two correlations with a mass term (which is needed, see \citealt{FalckeBiermann1996}), one arrives naturally at the fundamental plane.

One method to check if a correlation between two observables is only created by a third variable, is the partial correlation coefficient. In our case for radio and X-ray luminosity is the parameter that may create a spurious correlation the distance. The correlation coefficient is defined as:
\begin{equation}
r_{rx,d} = \frac{r_{rx} - r_{rd} r_{xd}}{\sqrt{(1.-r_{xd}^2) (1.- r_{rd}^2)}},
\end{equation}
where $r_{ab}$ is the normal Pearson R for a and b. This correlation coefficient should remove the effect of the different distances.

If we fix the mass coefficient $\xi_{\text M} = -0.85$, we find $r_{rx,d} = 0.91$. We have written a small Monte Carlo simulation, to test if this coefficient is significant. We assume a uniform distribution for the distances and the fluxes in log-space, and assume that the fluxes are uncorrelated. We created $10^6$ artificial datasets, and found that not a single dataset had a partial correlation coefficient as large as 0.8, even though the average normal correlation coefficient for the luminosities was 0.99. The standard deviation of the partial correlation coefficient is 0.11 and the mean, 0. Thus, the radio/X-ray correlation with a fixed mass coefficient is significant at the 8 $\sigma$ level. However, in our sample, we only include objects with measured radio and X-ray fluxes and do not include upper limits which could affect the significance.

On the other hand, if we fix $\xi_{\text M} = 1.4$ the fundamental plane suggests that $\frac{L_X}{L_R^{1.4}}$ and $M$ are correlated. The Kendall $\tau$ for this correlation is $\tau = 0.6$, which is significant as for uncorrelated data $\tau$ is normally distributed around 0 with a standard deviation of $0.006$. Again we can check whether this is due to the different distances in the sample: The partial correlation coefficient is $r_{L_X/L_R^{1.4},M,D} = 0.84$, which is again significant (7 $\sigma$). 

Even if the observational flux limits can not create a spurious fundamental plane, these limits might still bias the estimated parameters. This problem can be tested with our Monte Carlo simulation. We start with our fundamental plane and observe the parameter changes due to increasing flux limits. As the flux limits may reduce the observed scatter around the fundamental plane, we increase the intrinsic scatter by a factor of 2. The results are summarized in table \ref{tabMCTest}. As long as the flux limits stay in a reasonable range, the changes to the parameters are below 0.1, i.e., they are of the order of the uncertainties of the fits. 

\begin{table*}
\caption{Effect of the observing flux limits on correlated data: Most radio fluxes in our sample are obtained with the VLA, which can detect 0.1 mJy within a 10 minute snapshot. Chandra can go as deep as $10^{-14}$ erg/s/cm$^2$ in a 1000 s observation. \label{tabMCTest}} 
\begin{center}
\begin{tabular}{lccccc}
\hline \hline
\noalign{\smallskip}
Radio limit& X-ay limit &  $\xi_{\text R}$ & $\xi_{\text M}$ & $b_{\text X}$ \\
\hline \noalign{\smallskip}
0& 0 & $1.4 \pm 0.11$ &  $-0.86 \pm 0.15$ & $-4.84 \pm 3.2$ \\
0.5 & $10^{-13}$ & $1.37 \pm 0.11$ & $-0.80 \pm 0.15$ & $-4.21 \pm 3.2$ \\
5 & $10^{-13}$ & $1.34 \pm 0.10$ & $-0.77 \pm 0.13$ & $-3.67 \pm 3.1$ \\
0.5 & $10^{-12}$ & $1.36  \pm 0.12$ & $-0.76 \pm 0.16$ & $-3.94 \pm 3.4$ \\
5 & $10^{-12}$ &$1.27 \pm 0.10$ & $-0.64  \pm 0.12$ & $-1.75 \pm 2.8$ \\
\end{tabular}
\end{center}
\end{table*}

\subsection{Comparison of the KFC sample and the MHDM sample}
The correlation coefficient $\xi_{\text R}$ in the MHDM sample and its subsamples seems larger than those usually found in the KFC sample and a similar effect can be found for $\xi_{\text M}$. The fits of the KFC sample and its subsamples are in agreement with the value found for LH state XRBs ($\xi_{\text R} = 1.4$), while the deviations are larger for the MHDM sample. This may be seen as a hint that the MHDM sample is not simply a continuation of the LH state correlation for XRBs, but it may contain other effects like a different source of emission. 

The main statistical difference between the MHDM sample and the KFC sample is the smaller intrinsic scatter of the latter: $\sigma_{\text int} = 0.39$ compared to $\sigma_{\text int} = 0.65$. It is hard to assess the uncertainties of this value due to selection effects as the underlying distribution is unknown. Bootstrapping yields an error for both values around  $0.06$.

The discrepancy is partly due to the fact that the KFC sample has less AGN compared to XRBs than the MHDM sample. However, even if one adds the Seyferts and transition objects of \citet{NagarFalckeWilson2005} to create a sample of similar size than the MHDM sample, then the intrinsic scatter is still less $(\sigma_{\text int} = 0.46)$ than for the MHDM case. The same is true for the subsamples of similar size.

The most homogeneous subsample is the sample containing only LLAGN, LH state XRBs and Sgr A$^*$. Here we find $\sigma_{\text int} = 0.11$. This low value is not only due to the fact that we overestimated the errors of the XRBs, as the numerical value is below 0.1 for the LLAGN sample without XRBs as well. Thus, the correlation is extremely tight for the lowest luminosity objects. If we extend this sample to slightly higher accretion rates, i.e., include FR-I RGs and BL Lac objects the scatter increases due to peculiarities of these objects. The scatter further increases if we included objects FKM classify as high state objects. Thus, the reduced scatter supports the classification of AGN classes by FKM and suggests that there is a difference between LH state AGN and HS state objects. 

\subsection{Interpretation in the context of the proposed models}
In Fig.~\ref{Chi2map}, we show the $\hat{\chi}^2$ maps of the different samples and the predictions of the jet model and one disk/jet model. For the 'jet only' model the predicted values are: $\xi_{\text R} = 1.38$ and $\xi_{\text M} = -0.81$ (FKM)  while the values for the 'ADAF/jet' model are $\xi_{\text R} = 1.64$ and $\xi_{\text M} = -1.3$ (MHDM). The exact value for a disk/jet model depends on the solution used for the accretion flow. 

The full KFC sample contains some sources, for which we used optical fluxes to derive our equivalent X-ray luminosities. Thus, the 'ADAF/jet' does not have to be valid. If we ignore this, 
the full KFC sample seems to favour the 'jet only' model as, according to the confidence region, we can rule out the 'ADAF/jet' possibility by more than 3 $\sigma$. {\it However}, the exact fit values depend on the choice of the sample and the used assumptions for the statistical model. 

In contrast to the claims of \cite{Heinz2004}, the correlation {\it is} in agreement with a simple jet model in the regime where radiative cooling is not important. The choice of sources minimizes the effect of radiative cooling and the synchrotron cut-off as discussed in Sec.~\ref{SyncCut}, but it can not be ruled out that cooling has some effect on this sample.

The correlation with the least scatter is found for the subsample of the KFC sample containing only LH state XRBs, Sgr A$^*$ and LLAGN. Its confidence region is shown in the middle row of the figure. As this subsample consists of 'real' radio/X-ray data, both models claim to be valid. However, the fit is not in good agreement with both models. The 'jet only' model is disfavoured with $\approx 3 \sigma$ while the 'ADAF/jet' model is even stronger rejected. The remarkable low scatter in this subsample $\sigma_{\text Int} = 0.10$ supports the idea that LH state XRBs and LLAGN are indeed associated. 

In the light of the jet model, it may be that the deviation of the fit from the predicted value is due to synchrotron cooling and the synchrotron cut-off. Even though we have designed this sample to minimize their effect, it may still play a role in some of the objects. To analyze this effect, one would have to compare the X-ray spectrum with the assumed hard power-law and take the spectral index into account. If one finds significant deviations in the spectral index one will have to resort to a more complicated study where the measured spectral index or a more complicated model of the SED is taken into account in the fitting. 
Another possible explanation for this deviation, is that we do not treat the coupled errors of the XRB data points correctly. As we treat all errors as independent we do not constrain the correlation index $\xi_{\text R}$ from the XRBs as well as we could. If we set this index to  $\xi_{\text R}=1.4$ and only fit $\xi_{\text M}$ we find $\xi_{\text M} = -0.85$ for the KFC subsample containing LLAGN and XRBs. See also Fig.~\ref{SetXiR}. This is roughly in agreement with the value predicted by the jet/synchrotron model of $\xi_{\text M} = -0.81$. 

For a 'disk/jet' model the discrepancy may be due to the disk model used. Besides the discussed ADAF solution, one can use any other accretion flow model to create the X-rays. This can change the prediction considerably, see e.g., MHDM. Thus, the discrepancy of the fit compared to the model predictions can not rule out any of the suggested models with certainty, but -- in contrast to earlier claims -- it does not support them either.

To compare the edited MHDM sample with the models, we now have to note
that the 'jet only' model claims to become invalid for the X-ray
emission of high state objects like Quasars. Interestingly, the edited
MHDM sample including high-state objects indeed disfavours the 'jet
only' model and is in agreement with the 'ADAF/jet' model (1.2
$\sigma$). However, also for this sample the selection effects are
dominating the exact fit value as well.

\subsection{The conspiracy}
We have seen that the fundamental plane in the two described incarnations is only slightly different. For both samples we find only slightly different parameters, and the scatter seems to be less in the case of the KFC sample. The MHDM sample contains low luminosity objects as well as bright quasars. However, there do not seem to be obvious outliers.

The radio emission is usually attributed to the jet. For the higher observation frequencies there are objects in the two samples that are clearly jet dominated and others for which the accretion flow will be the dominant part at higher frequencies. The clearest examples for synchrotron emission are the BL Lac objects in the KFC sample but, also for FR-I RG this origin is well established \citep{ChiabergeCapettiCelotti1999}. On the other hand, the X-ray emission from radio quiet Quasars is very likely not synchrotron emission. Nevertheless, even though the inclusion of the Quasar subsample changes the correlation  and increases the scatter, they do not drop off the correlation like HS state XRBs. Likewise, if the X-rays of LH state XRBs and LLAGN are created by the accretion flow, why do the BL Lac objects and FR-I RGs still follow the fundamental plane? There seem to be a "fundamental plane conspiracy": even though the emission processes are different the objects all lie near the fundamental plane.

Within the jet model one can explain part of the conspiracy by the different emission processes (see Fig.~\ref{fiJetSpec}). If we observe a source at a frequency after the synchrotron cut-off, the Compton branch takes over. The inverse Compton emission can in many sources reach the values one would find if one extrapolates the synchrotron power-law to X-ray frequencies. It will mainly increase the scatter in the correlation.

\section{Conclusions}
In the previous sections we have reconfirmed the existence of the fundamental plane of accreting black hole in the black hole mass, radio and X-ray luminosity space.
We find that the result of a statistical analysis of the radio/X-ray correlation depends strongly on the assumptions of the distribution and magnitude of the measurement errors and the intrinsic scatter. The measurement uncertainties have been taken from the literature. The unknown intrinsic scatter, e.g., the scatter due to relativistic beaming, non-simultaneous observations or source peculiarities, has been parameterized and estimated for the observed samples. 

Using this refined method we compared the proposed radio/X-ray correlations of MHDM and the improved KFC sample based on FKM. Both samples differ in their source selection: while the KFC sample tries to include only sources belonging to the low/hard state, the MHDM sample includes all kinds of AGN. Also the observing frequencies differ for some sources, as the KFC sample uses extrapolated optical fluxes for FR-I RGs and BL Lac objects.  

The best fit values of both samples depend on the relative number of sources in each class of objects, e.g., the relative number of quasars or FR-I RGs compared to LLAGN. This can be understood if different physical processes are dominant in the different classes, e.g., the emission from LLAGN may be due to the jet, while for quasars the disk emission dominates. 
The confidence regions do not reflect this problem and have to be viewed as a lower bound on the errors of the parameters.

The best fit values found for the KFC sample are $\xi_{\text R} = 1.41 \pm 0.11 $ and $\xi_{\text M} = -0.87 \pm 0.14$ while we find for the MHDM sample  $\xi_{\text R} = 1.74 \pm 0.20$ and $\xi_{\text M} = -1.35 \pm 0.27$. Thus, the KFC sample suggests a simple uncooled 'jet only' model while the MHDM sample favours the 'ADAF/jet' model. However, the selection effects are very hard to control.

The KFC sample seems to be a more homogeneous sample, as it has a
lower intrinsic scatter. The fundamental plane for the subsample
containing only LLAGNs and XRBs is surprisingly tight with a scatter
of $\sigma_{\text int} = 0.12$ dex, while the full sample has
$\sigma_{\text int} = 0.38$. Compared to this, the MHDM sample has a
higher intrinsic scatter of $\sigma_{\text int} \approx 0.6$ dex. This
supports the AGN classification of FKM in low/hard and high/soft state objects. 

In general, the fundamental plane of black hole activity is confirmed
by our analysis. With a careful control of a homogeneous source
selection (high-state versus low-state), the scatter can reach rather
low values. This promises a wider application of the ``fundamental
plane'' in other contexts (see e.g., \citealt{Merloni2004,Maccarone2005}) and calls for improved radio and X-ray surveys in the future. 

{\it Acknowledgements}
The authors thank Elena Gallo for the provision of the LH state XRB data points in electronic form. We thank Sera Markoff, Tom Maccarone, Sebastian Jester and Matthias Kadler for helpful discussions. We thank our referee for constructive comments. 

\bibliography{refs} 
\bibliographystyle{aa}

\end{document}